# Anomalously large oxygen-ordering contribution to the thermal expansion of untwinned $YBa_2Cu_3O_{6.95}$ single crystals: a glass-like transition near room temperature

Peter Nagel, Volker Pasler and Christoph Meingast
*Forschungszentrum Karlsruhe, Institut für Festkörperphysik, PO Box 3640, 76021 Karlsruhe, Germany*
and
Alexandre I. Rykov and Setsuko Tajima
*Superconductivity Research Laboratory-ISTEC, 10-13 Shinonome 1-Chome, Koto-ku, Tokyo 135, Japan.*

We present high-resolution capacitance dilatometry studies from 5 - 500 K of untwinned $YBa_2Cu_3O_x$ (Y123) single crystals for $x \sim 6.95$ and $x = 7.0$. Large contributions to the thermal expansivities due to O-ordering are found for $x \sim 6.95$, which disappear below a kinetic glass-like transition near room temperature. The kinetics at this glass transition is governed by an energy barrier of $0.98 \pm 0.07$ eV, in very good agreement with other O-ordering studies. Using thermodynamic arguments, we show that O-ordering in the Y123 system is particularly sensitive to uniaxial pressure (stress) along the chain axis and that the lack of well-ordered chains in Nd123 and La123 is most likely a consequence of a chemical-pressure effect.

The superconducting and normal state properties of $YBa_2Cu_3O_x$ (Y123) depend strongly both on the total oxygen content (O-content) x and on the degree of O-order in the CuO chain layer, and a great deal of experimental and theoretical effort has gone into studying both the different types of O-ordered structures and their correlation with superconducting properties [1-23]. Most of these studies have concentrated on samples with O-contents in the range $6.3 < x < 6.7$, probably because superconductivity is particularly sensitive to the degree of O-order in this range. The kinetics of O-rearrangements occurs via a thermally activated diffusion process with a time constant on the order of hours at room temperature, and many interesting effects have been observed by room-temperature annealing studies [7-13].

There are however only a few studies in which O-ordering effects have been reported near the optimal doping level ($6.9<x<7.0$), where there are only a few O-vacancies and much smaller ordering effects are expected. Nevertheless, resistivity studies of Lavrov et al. clearly demonstrated that O-ordering occurs even in nearly fully doped ($x \sim 6.95$) samples [14]. Further evidence for O-ordering in optimally doped crystals has also been obtained by high-temperature quench experiments of Erb et al., in which it was clearly shown that O-vacancy clusters form in slowly cooled crystals and that these clusters form strong flux-pinning centers, which are responsible for the 'fish-tail' effect [16].

In this Letter we present a thermodynamic and kinetic study of O-ordering processes in near optimally doped ($x \sim 6.95$) Y123 untwinned single crystals using high-resolution thermal expansion measurements. Surprisingly large contributions to the linear thermal expansion coefficients (expansivities) from O-ordering are observed. These are, as expected, absent in the fully oxygenated ($x=7.0$) state. The large anomalies imply that O-ordering is highly sensitive to slight dimensional changes of the crystal lattice. The largest contribution occurs along the chain axis (b-axis), from which the peculiar result, that compressing the chain axis increases the degree of O-order, can be inferred. Our results clearly show that the lack of well-ordered chains in Nd123 and La123 [17] is a consequence mainly of the larger (in comparison to Y123) unit cell.

Our investigations have been performed on Y123 single crystals grown by a pulling technique, detwinned at a uniaxial pressure of 10 MPa and oxygenated at 490°C [24]. This procedure results in a slightly overdoped sample with an O-content x of approximately 6.95. In order to get to the fully oxygenated state with $x = 7.0$, the crystals were annealed at 376 bar O-pressure and 400°C for 140 hours. The same crystal, with linear dimensions of $6 \times 3 \times 2$ mm$^3$ for the different orthorhombic axes a, b and c, respectively, as used in a previous thermal expansion investigation [25] was used presently. The thermal expansion was measured with two different high-resolution capacitance dilatometers in the range 5 - 300 K and 150 - 500 K respectively, using constant heating and cooling rates between 2.5 and 20 mK/s. Data points were taken every 100 mK.

In Fig.1 we present the linear thermal expansivities, $\alpha(T) = \frac{1}{L} \cdot \frac{dL}{dT}$, of Y123 along the three orthorhombic axes in the temperature range 5 - 500 K for two different x values [$x \sim 6.95$ (solid line) and $x = 7.0$ (dashed line)]. In addition to the well-studied superconducting anomalies at $T_c$ [25-27], a quite large and broad $\alpha(T)$ anomaly is observed for the $x \sim 6.95$ sample near room temperature, or more precisely at $T_g = 280$ K, along each of the axes. No comparable effect is seen for $x = 7.0$, and in the following we show that this anomaly results from a kinetic glass-like transition due to O-ordering.

The shape of this anomaly (width [28] and hysteresis for heating and cooling curves) are typical of kinetic 'glass' transitions [29-31]. A comparison between the $x \sim 6.95$ and $x = 7.0$ data sets clearly indicates that, as expected [29-31],

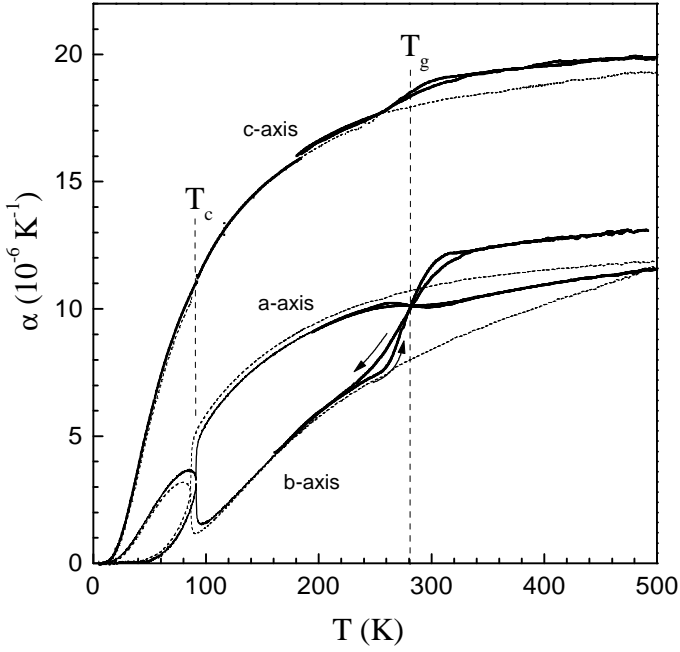

**Fig.1.** Expansivity versus temperature of $YBa_2Cu_3O_x$ for the a, b and c axes for x ~ 6.95 (solid lines) and x = 7.0 (dashed lines). In addition to the superconducting phase transition at $T_c$, a glass-like transition can be observed at $T_g$ in the data with x ~ 6.95. The arrows indicate heating and cooling cycles.

there are additional contributions ($\Delta\alpha$) to the expansivities above $T_g$, which are absent below $T_g$. The contribution above $T_g$ is positive for the b and c-axis, and negative for the a-axis [32] (see Fig. 1). The largest effect is found for the b-axis, for which we have also examined the expansivity as a function of cooling rate for -2.5, -5, -10 and -20 mK/s. As seen in Fig. 2, the glass transition shifts to higher temperatures with increasing cooling rates, which clearly indicates the kinetic nature of the transition. Here we have subtracted a background, which was constructed by shifting the data of the b-axis for x = 7 by a small value, from the original data. In order to extract a well-defined $T_g$ from these broad transitions, we have normalized the data in Fig. 2 by the equilibrium expansivity $\Delta\alpha_b^{equil.}$ (dashed line). Below $T_g$, $\Delta\alpha_b^{equil.}$ is obtained by an extrapolation of the high-temperature data to lower temperatures. In Fig. 3 these normalized curves are plotted versus $-T^{-1}$ in order to show that the glass transition shifts in temperature, while the shape of the glass-transition does not change. $T_g$ values can now easily be extracted from Fig. 3 as the midpoints of the transitions, and these are shown in an Arrhenius plot (see inset in Fig. 3). The data show a good linear dependence, indicative for a thermally activated relaxation time [29-31]

$$\tau(T) = \nu_0^{-1} \cdot e^{\frac{E_a}{k_B T}} \quad (1)$$

with an activation energy $E_a = 0.98 \pm 0.07$ eV. The relaxation time $\tau$ at $T_g$

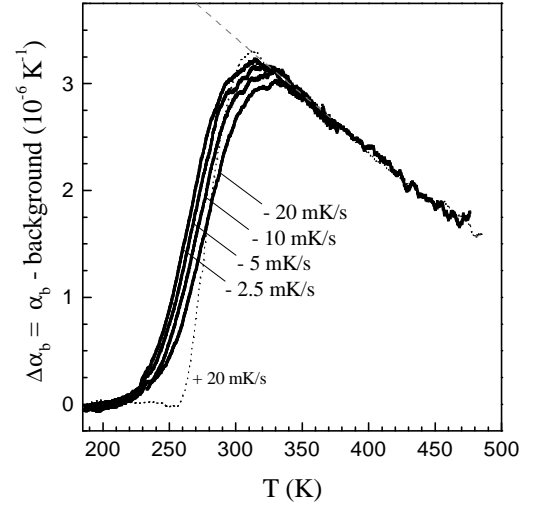

**Fig.2.** Glass-like contribution to $\alpha_b(T)$ of $YBa_2Cu_3O_{6.95}$ for different cooling rates (solid lines) and heating with 20 mK/s (dotted line) after cooling with identical rate. The dashed line represents $\Delta\alpha_b^{equil.}$ (see text).

$$\tau(T_g) = \frac{k_B \cdot T_g}{E_a} \cdot \frac{T_g}{|r|} \quad (2)$$

can be directly related to the cooling rate, r, in our experiment [33], yielding the pre-factor $\nu_0 = 1 \cdot 10^{15}$ Hz in Eq. 1. Our value for the activation energy $E_a$, which is physically related to the barrier for O-hopping, is in very good agreement with those found by other kinetic studies of O-ordering [5,8,15,34]. This and the fact, that the glass transition vanishes for x = 7, provides clear evidence, that the expansivity anomalies in the x ~ 6.95 sample result from the freezing of O-ordering processes. This ordering occurs on a 'local' scale, as indicated by the short extrapolated diffusion lengths using the experimental diffusion constants [34] and our relaxation time $\tau(280\ K) = 350$ s. We stress that we see no indications for the so-called '240 K - phase transition', which has also been discussed in connection with O-ordering [35]. In the following, we analyze the thermodynamics of the O-ordering above $T_g$, where thermal equilibrium is guaranteed.

The additional contribution to the expansivity $\Delta\alpha$ above $T_g$ (see Fig. 2) is a direct measure of the uniaxial pressure (stress) dependence of the degree of O-order at constant temperature [36]

$$\Delta\alpha_i = -\frac{1}{V_{mol}} \cdot \left.\frac{\partial S_{Oxygen}}{\partial p_i}\right|_T \quad (i = a, b, c). \quad (3)$$

Here, $S_{Oxygen}$ is the molar entropy associated with O-disorder and $V_{mol}$ is the molar volume. Inserting the measured $\Delta\alpha$ values at 340 K ($\Delta\alpha_a = -5 \cdot 10^{-7} K^{-1}$, $\Delta\alpha_b = 30 \cdot 10^{-7} K^{-1}$, $\Delta\alpha_c = 8 \cdot 10^{-7} K^{-1}$) into Eq. 3, we obtain $dS_{Oxygen}/dp_a = 56$ mJ/(mol·K·GPa), $dS_{Oxygen}/dp_b = -311$ mJ/(mol·K·GPa) and $dS_{Oxygen}/dp_c = -85$ mJ/(mol·K·GPa). This shows that uniaxial pressure (stress) along the a-axis decreases the

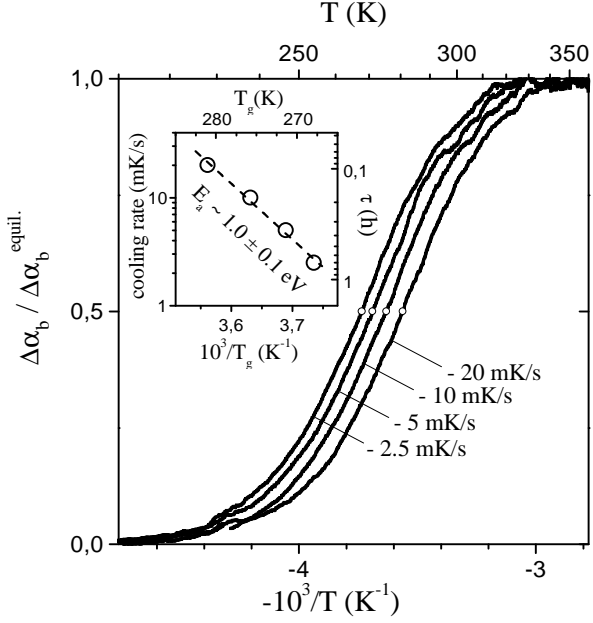

**Fig.3.** Normalized plot of $\Delta\alpha_b(T)$ versus $-T^{-1}$ for x ~ 6.95. The inset shows an Arrhenius-plot of the cooling rate and relaxation time versus $T_g^{-1}$ (see text for details).

O-order, whereas uniaxial pressure along the b- and c-axis increases the O-order. Hydrostatic pressure is expected to increase the O-order since $dS_{Oxygen}/dp_{hyd} = \Sigma\, dS_{Oxygen}/dp_i =$ -0.34 J/(mol·K·GPa). We can also calculate the change of $S_{Oxygen}$ resulting from volume changes

$$\frac{\Delta S_{Oxygen}}{\Delta V_{mol}/V_{mol}} = B \cdot \Delta\beta \cdot V_{mol} = \frac{0.34 \frac{J}{mol\cdot K}}{\%\text{ volume change}}, \quad (4)$$

using the bulk modulus B = 100 GPa [37] and the O-ordering contribution to the volume expansivity $\Delta\beta$. In order to get a feeling for the magnitude of these pressure (volume) effects, it is instructive to calculate the maximum entropy difference $\Delta S_{Oxygen}^{max}$ between totally ordered and totally disordered states. For x = 6.95 (5 % O-vacancies) and assuming that the disorder is limited to the disorder in the chains (no O-5 site occupations), $\Delta S_{Oxygen}^{max}$ can be easily calculated using elementary statistical mechanics to equal 1.65 J/(mol·K) [38]. Thus, a relatively small (1 %) volume increase (decrease) will result in a quite large (20 % of $\Delta S_{Oxygen}^{max}$) decrease (increase) of $S_{Oxygen}$. This result, which is a direct thermodynamic consequence of our data, provides a simple explanation for the different degrees of O-order observed in R123 (R= rare-earth) compounds [17]. In R123 compounds with a larger (smaller) unit cell (in comparison to Y123) the degree of O-order should be much lower (much higher) than in Y123. Nd123 e.g. has a unit-cell which is about 2.6 % larger than that of Y123 [39] and should therefore have an entropy, which is about 0.88 J/mol·K larger than in Y123 - this entropy-difference is about 50 % of $\Delta S_{Oxygen}^{max}$, and suggests that Nd123 is nearly fully disordered, in agreement with experiment [17]. Also, the failure to detect any glass-transition in Nd123 with x ~ 6.95 up to 500 K using dilatometry [18], which implies that dS/dV = 0 (see Eq. 4), suggests that $S_{Oxygen}$ is at its maximum value. Yb123, on the other hand, has a smaller unit cell and is expected to have a higher degree of chain order. Our results suggest that the difference between the degree of O-ordering in the R-123 can be explained by a simple chemical-pressure effect.

The additional contribution $\Delta\alpha_b$ (see fig.2) and, thus, $dS/dp_b$ decreases strongly in magnitude with increasing temperature, and this provides additional thermodynamic information about the O-ordering. An analogous contribution has to also occur in the specific heat $C_p$, but, unfortunately, even high-resolution measurements have not detected this anomaly [40]. These specific heat data, however, provide an upper limit of about 2 J/(mol·K) for this contribution at 300 K. It is instructive to compare this value with the difference $\Delta C$ between the specific heats at constant volume $C_V$ and constant pressure $C_p$, which can be directly calculated from our volume expansivity $\beta$ using

$$\Delta C = C_p - C_V = T \cdot V_{mol} \cdot B \cdot \beta^2. \quad (5)$$

Just above $T_g$, $\Delta C$ = 0.8 J/(mol·K), which is about 40 % of the maximally estimated $C_p$ contribution. This indicates that, in addition to the usual increase in entropy due to increasing temperature, the volume change due to the thermal expansion plays a major, if not dominating, role in the increase of entropy with increasing temperature; this is because the dS/dV term is so large. Another way to quantify the strong volume dependence is through a partial Grüneisen parameter for O-ordering

$$\gamma_{Grüneisen}^{O-ordering} = \frac{\Delta\beta \cdot V_{mol} \cdot B}{\Delta C_p}. \quad (6)$$

Inserting the maximum estimate for the specific heat anomaly just above $T_g$ into Eq. 6 yields a lower limit for $\gamma_{Grüneisen}$ (> 17). An upper limit ($\gamma_{Grüneisen}$ < 43), on the other hand, is obtained by realizing that $\Delta C_v$ must be greater than or equal to zero, from which a lower limit of $\Delta C_p \geq$ 0.8 J/(mol·K) can be inferred (see Eq. 5). Such large values of $\gamma_{Grüneisen}$ clearly document the enormous pressure/volume dependence of the O-ordering effects in Y123.

So far we have not discussed either the spatial arrangements or the microscopic origin of these ordering processes. From optical measurements it is clear that the chain length is growing as one lowers the temperature [19]. This is also compatible with the hole transfer from chains to planes, which is expected as the chains increase in length [20], observed by Lavrov at al. [14], as well as with the formation of O-clusters [16]. Theoretically, O-ordering has been modeled using the asymmetric-next-nearest-neighbor-interaction (ASYNNNI) model, in which several phenomenological parameters (V1 > V2 > V3 > ...) are introduced to describe the O-interactions in the basal plane

[21-23]. We note that we do not observe the thermodynamic phase-separation line (ortho-I ↔ ortho-I + ortho-II) predicted by this model [22] in our experimental temperature window (280 K < T < 500 K). The parameter V2 ≈ 2000 K determines the length of the chains in this model. A recent theoretical study has argued that a single V2 parameter is too simple, and that this interaction depends on chain length as well as on x [41]. Our results show that V2 depends also strongly on the size of all lattice parameters, which demonstrates the complexity of this interaction. We should point out that this cannot be understood in terms of a simple steric effect involving just the b-axis parameter, since V2 also depends on the size of the other axes, and any ab-initio calculations of V2 will most likely have to include the electronic structure of the whole unit cell, including effects such as charge transfer etc.

In conclusion, our thermal expansion data of untwinned single crystals clearly reveal the kinetics and, for the first time, also the thermodynamics of O-ordering effects in $YBa_2Cu_3O_{6.95}$. These ordering effects manifest themselves in terms of a kinetic glass-like transition near room temperature. The large size of the glass-transition anomalies along all axes are directly correlated with the extreme dependence of the degree of O-order with the lattice parameters, and this provides a natural explanation of the different O-order (e.g.) in Nd123 and La123 in terms of a simple chemical-pressure effect. Studies of O-ordering in R123 compounds using high-pressure or strain induced by lattice mismatch in thin films should provide further valuable information [42]. Theoretical modeling of the O-ordering via lattice gas models such as the ASYNNNI model, should incorporate this new results for a more quantitative comparison with experiments.

This work was partially supported by NEDO, Japan. We would like to thank A. Erb, H. Claus and T. Wolf for useful discussion.